\documentclass[a4paper, 11pt]{article}

\usepackage{verbatim}
\usepackage{comment}
\usepackage{url}
\usepackage{bbm}
\usepackage{paralist}
\usepackage{color}
\usepackage[T1]{fontenc}
\usepackage[utf8]{inputenc}
\usepackage{authblk}
\usepackage{natbib}
\usepackage{amsmath}
\usepackage{graphicx}
\usepackage[margin=2.5cm]{geometry}

\newcommand{\bs}[1]{\boldsymbol{#1}}
\newcommand{\mc}[1]{\mathcal{#1}}
\newcommand{\real}{\mathbbm{R}}

\title{Bayesian regional flood frequency analysis for large catchments}

\author[1]{Thordis L. Thorarinsdottir}
\author[1]{Kristoffer H. Hellton}
\author[1]{Gunnhildur H. Steinbakk}
\author[2]{Lena Schlichting}
\author[2]{Kolbj{\o}rn Engeland}

\affil[1]{Norwegian Computing Center, Oslo, Norway}
\affil[2]{Norwegian Water Resources and Energy Directorate (NVE), Hydrological Modeling Section, Oslo, Norway}

\begin{document}

\maketitle

\begin{abstract}
Regional flood frequency analysis is commonly applied in situations where there exists insufficient data at a location for a reliable estimation of flood quantiles. We develop a Bayesian hierarchical modeling framework for a regional analysis of data from 203 large catchments in Norway with the generalized extreme value (GEV) distribution as the underlying model. Generalized linear models on the parameters of the GEV distribution are able to incorporate location-specific geographic and meteorological information and thereby accommodate these effects on the flood quantiles. A Bayesian model averaging component additionally assesses model uncertainty in the effect of the proposed covariates. The resulting regional model is seen to give substantially better predictive performance than the regional model currently used in Norway. 
\end{abstract}

\section{Introduction}
Flood frequency analysis (FFA) is a statistical data-based approach to determine the magnitude of a flood event with a certain return period. If sufficient data are available at a single site, an extreme value distribution is usually fitted to the observed data (at-site FFA or local model). However, ungauged sites or sites suffering from incomplete data require the use of data from nearby or comparable gauged stations. We refer to this as regional flood frequency analysis (RFFA).

The motivation for this study is the need to update the current guidelines for estimation of the design flood in ungauged catchments in Norway \citep{Castellarin&2012, Midttomme&2011}. The current guidelines are based on recommendations that are more than 20 years old \citep{Saelthun&1997}. With increased data availability (20 more years) and the development of new Bayesian inference methods, including for engineering applications (e.g. \citet{Ball&2016}), there is a significant potential for improvement to the current RFFA methods. 

The classical approach for RFFA is the index-flood method, see e.g. \citet{Dalrymple1960} and \citet{HoskingWallis1997}. This method assumes that the flood frequency curve for all sites in a region follows the same distribution up to a scaling factor. The index flood method hence consists of three independent steps: (1) identification of homogeneous regions or similar stations, (2) estimation of the index flood, and (3) derivation of the growth curve that gives factors for scaling an index flood to a suitable return level. If no appropriate data are available, the index flood is derived by regional regression analysis or based on nearby measurement stations (scaled with respect to catchment area), and a regional growth curve is applied.  An overview of European procedures for RFFA is given in \citet{Castellarin&2012}. For step (1), typical approaches are to use fixed geographical regions, station similarity, or focused pooling (region of influence). In Austria, and interpolation approach is used. For step (2), linear regression, possibly combined with transformation of independent and/or dependent variables is used. For step (3), maximum likelihood, ordinary moment estimation or estimation based on l-moments are commonly used. 

The regional approach currently used in Norway is based on (1) fixed geographical regions, (2) linear regression on transformed variables, and (3) a unique growth curve for each region. The growth curve of each region is an average of all growth curves within a region based on an l-moment estimator. This approach does not account for parameter uncertainty, neither in the regression equation for the index flood estimation nor in the growth curves. The model uncertainty originates from the model selection in the regression analysis for the index flood model as well as the selection of a parametric distribution for the growth curve, whereas the parameter uncertainty originates from a limited sample size and measurement errors \citep[e.g.][]{Steinbakk&2016}.

A Bayesian approach accounts for these uncertainties and easily provides the predictive distribution of design floods \cite[e.g.][]{FawcettWalshaw2016, Kochanek&2014}. The potential for increasing the reliability of design flood estimates by including/improving the knowledge basis from which flood estimates are derived has increased the popularity of Bayesian methods. In addition, such methods are well suited for combining sources of information, such as historical information and expert judgments \citep[e.g.][]{ParentBernier2002, ReisStedinger2005}. In the most recent update of the guidelines for FFA in Australia, the Bayesian approach is recommended \citep{Ball&2016}. In a recent study on RFFA in small catchments in Norway, a Bayesian index flood approach is recommended \citep{Glad&2014}.   

Bayesian hierarchical models taking into account spatial and temporal structures have been used to describe extreme values in meteorology including extreme precipitation \citep{Cooleyetal2007, Renard&2013, Dyrrdal&2015}, wind and storm surges \citep{FawcettWalshaw2016}. These Bayesian hierarchical models have a data layer described by an extreme value distribution at each site which depends on some unknown parameters, typically location, scale and shape. The spatial dependency is, in most cases, modeled by letting each parameter in the extreme value distribution depend on geographical, meteorological, and other location-specific characteristics. For some applications, a spatial dependency is based on distance, e.g. the Austrian RFFA procedure described in \citet{Castellarin&2012} and the regional approaches for precipitation presented in \citet{Renard&2013} and \citet{Dyrrdal&2015}.  

The main objective of this study is to establish a hierarchical Bayesian model that can be used for estimating design floods in ungauged catchments. The following sub-objectives were identified:
\begin{compactitem}
\item Estimate the regional model and identify the most important predictors;
\item Evaluate the predictive performance of the regional model;
\item Compare the regional model to the existing model of \citet{Saelthun&1997}.
\end{compactitem}
In order to achieve these aims, we build upon the work of \cite{Dyrrdal&2015} and use a hierarchical Bayesian model to spatially describe the parameters of a generalized extreme value (GEV) distribution for annual maximum daily discharge in 203 catchments in Norway.  For comparison, we also fit a local extreme value distribution to the discharge series at individual sites, using a three-parameter GEV distribution if more than 50 years of data are available and the simpler two parameter Gumbel distribution if between 20 and 50 years of data are available, following the current guidelines for statistical FFA in Norway \citep{Midttomme&2011}. 

The remainder of the paper is organized as follows: Section 2 and 3 present the data and the regional GEV model, detailing the Bayesian hierarchical framework and Markov sampling, respectively. Section 4 gives an overview of the regional model currently used in Norway, and Section 5 outlines the setup for model validation. Section 6 presents first the resulting model and discusses model validation in terms of reliability and stability, then explores individual return levels for certain stations, and ends with a comparison of the model to the current regional and local models. The final Section 6 contains some concluding discussion and details of the Bayesian inference algorithm are given in the appendix.  

\section{Data}
\begin{figure}[!hbpt]
\centering
\includegraphics[width=\textwidth]{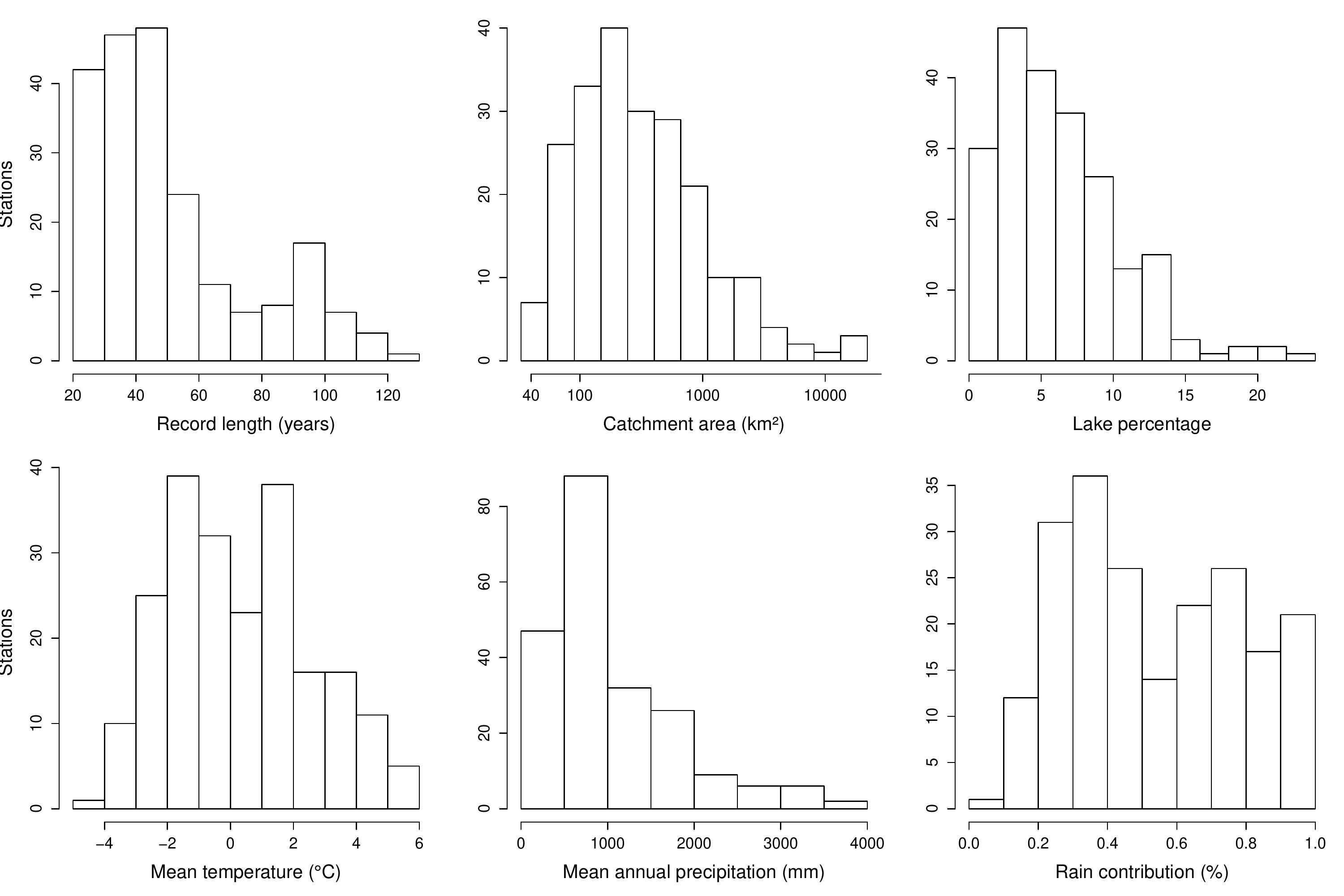}
\caption{Histograms for record length (years), catchment areas ($\text{km}^{2}$), lake percentage ($\%$), mean annual temperature ($^{\circ}$C), mean annual precipitation (mm) and the estimated rain contribution to floods ($\%$). 
}\label{fig:histCovar}
\end{figure}

\begin{figure}[!hbpt]
\centering
\includegraphics[width=1.1\textwidth]{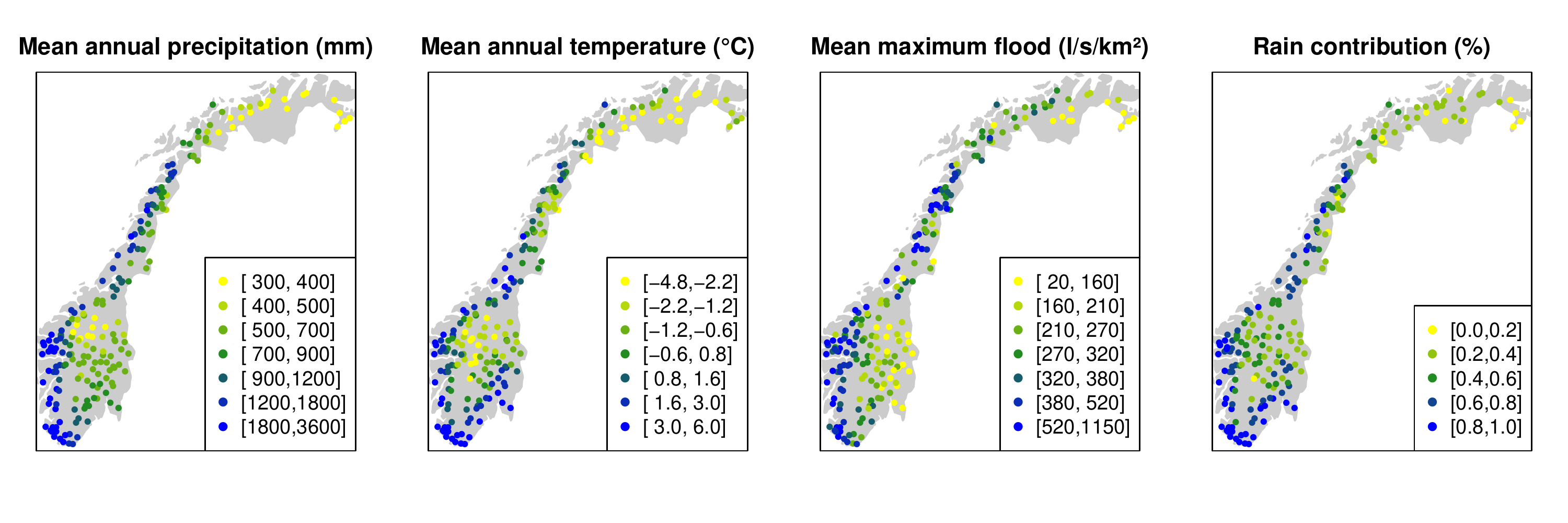}
\caption{Spatial distribution of mean annual precipitation (mm), mean annual temperature ($^{\circ}$C), mean annual maximum floods ($\text{l/s/km}^{2}$) and the rain contribution to floods ($\%$).}\label{fig:MAPcovar}
\end{figure}

The flood data consist of annual maximum floods from 203 streamflow stations of the Norwegian hydrological database ``Hydra II'' with at least 20 years of quality controlled data for periods with minimal influence from river regulations (see \citet{Engeland&2016} for details). For all gauging stations, we extracted a set of catchment properties as listed in Table~\ref{tab:covariates}. Climatological temperature and precipitation information are derived from the $1 \times 1$ km daily data product SeNorge available at \url{www.senorge.com}. Histograms for record length, catchment areas, lake percentage, mean annual temperature and precipitation, and the contribution of rain to floods is shown in Figure \ref{fig:histCovar}. Figure \ref{fig:MAPcovar} shows the spatial distribution of mean annual precipitation and temperature, mean annual maximum floods and rain contribution. 

\begin{table}[!hbpt]
\centering
\caption{Overview of covariate information used in the regional Bayesian model.} \label{tab:covariates}
\begin{tabular}{l p{10cm}} 
\hline
Explanatory variables  & Description\\ \hline
Longitude  & \\
Latitude & \\
Effective lake  & Percent of effective lake \\
Average fraction of rain & Average relative contribution of rain (vs. snowmelt) in the floods  \\
Catchment area & Total area of catchment, also including parts outside Norway\\
Inflow & Average inflow per year\\
Average rain in April  & Average over the period 1960--1990 \\
Average rain in August  & Average over the period 1960--1990\\
Snow melting in March  & Average over the period 1960--1990\\
Catchment gradient & Difference in  height meters between the 20th and 90th percentile of the gradient profile, standardized by the total catchment length \\
Exposed bedrock & Percent of mountainous area \\
Relative catchment area & Total area divided by catchment length \\
\hline
\end{tabular}
\end{table}

The catchment areas range between 50 and 18~110 km$^2$ with a median of 247 km$^2$. The presence of lakes influences flood sizes, and 49.1\% of the catchments have more than 1\% of the catchment area (effectively) covered by lakes. For these catchments, the median effective lake percentage is 2.73\%. The mean annual precipitation ranges from 291  to 3571 mm with a median of 732 mm. We see a strong west-east gradient in the spatial distribution with the highest precipitation on the west coast. The mean annual temperature ranges from -4.6 to 6.0 $^\circ$C with a median of 0.1 $^\circ$C. The temperature is influenced by elevation as well as latitude in that it decreases with both elevation and latitude. The relative contribution of rain is estimated by calculating the ratio of accumulated rain and snowmelt in a time window prior to each flood and then averaging these ratios for all floods, see \citet{Engeland&2016} for details. Rainfall gives the major contribution to floods in most coastal catchments, whereas snowmelt is important in inland, northern and high altitude catchments. The typical flood season for catchment dominated by snowmelt floods is spring and early summer while no clear seasonal patterns are seen for the catchments dominated by rain floods.

The flood records and the associated catchment properties (catchment area, record length, mean annual runoff and several other catchment descriptors) are available as supplementary materials. 

\section{Methods}
Extreme value theory provides a framework for modeling the tail of probability distributions. Let $V_1, \ldots, V_n$ denote continuous, univariate random variables assumed to be independent and identically distributed. If the normalized distribution of the maximum $\max\{V_1,\ldots, V_n\}$ converges as $n \rightarrow \infty$, then it converges to a generalized extreme value (GEV) distribution \citep{FisherTippett1928, Jenkinson1955}. For this reason, a GEV distribution is commonly used to model block maxima (the maxima over equally sized blocks of data) such as the annual maximum.  See \citet{Coles2001} for an introduction to the statistical application of extreme value theory.

In Norway, the three-parameter GEV distribution or a special case thereof, the two-parameter Gumbel distribution, are recommended for analyzing long data series from individual stations as these models have been found to provide the best fit for Norwegian data \citep{Midttomme&2011, Castellarin&2012}.  We have thus chosen to base our regional model on the three-parameter GEV distribution.  Alternative models may provide a better fit in other regions. An overview of methods that are used for operational flood frequency analysis in Europe is given in \citet{Castellarin&2012}. 

\subsection{Regional GEV model}
\subsubsection{Model formulation}
The data are given by series of annual maximum floods. Denote by $Y_{ts}$ the maximum flood in year $t \in \{1,\ldots, n_s\}$ at station $s \in \mc{S}$, the set of all stations in the data set, where $n_{s}$ is the number of annual floods observed at station $s$. We assume the floods to be independent and identically distributed over time with site specific covariates
\[
Y_{ts} \sim GEV(\mu_s,\kappa_s,\xi_s),\quad t \in \{ 1, \dots, n_{s}\}; s \in \mc{S}.
\]
The three parameter GEV distribution is here parametrized in terms of the location $\mu_s \in \real$, inverse scale $\kappa_s \in \real_+$, and shape $\xi_s \in \real$. The distribution is usually parametrized with the scale parameter $\sigma_s^2 = 1 / \kappa_s$, rather than the inverse scale \citep{Coles2001}, but the current parametrization is common in the Bayesian setting, see e.g. \citet{Dyrrdal&2015}. The density is given by 
\begin{equation}\label{eq:gev density}
pr_{GEV}(y_{ts}|\mu_s, \xi_s, \kappa_s) = \kappa_s h(y_{ts})^{-(\xi_s + 1)/\xi_s}\exp\left\{-h(y_{ts})^{-\xi_s^{-1}}\right\},
\end{equation}
for $h(y_{ts}) > 0$ with 
\[
h(y_{ts}) = 1 + \xi_s\kappa_s (y_{ts} - \mu_{ts})
\]
when $\xi \neq 0$.  For $\xi = 0$ the density is given by the Gumbel distribution
\[
pr(y_{ts}|\mu_s, \kappa_s) = \kappa_s h(y_{ts}) \exp\left\{-h(y_{ts})\right\}
\]
for $h(y_{ts}) > 0$ with 
\[
h(y_{ts}) = \exp(-\kappa_s (y_{ts} - \mu_{ts})).
\]
For the purposes of dam safety, we are interested in estimates of certain high quantiles of the resulting GEV distribution at site $s$. The tail behavior is driven by the value of the shape parameter $\xi_s$ and generally falls in three classes: The Fr\'echet type ($\xi_s > 0$) has a heavy upper tail, the Gumbel type ($\xi_s \rightarrow 0$) is characterized by a light upper tail, while the Weibull type ($\xi_s < 0$) is bounded from above. The shape parameter thus provides vital information on the statistical properties of the variable of interest and is, concurrently, difficult to estimate due to the involved parametric form of the density in \eqref{eq:gev density} as a function of $\xi_s$.  To estimate the quantile $p$ of the resulting GEV distribution function, we employ the GEV quantile function
\begin{equation}\label{eq:GEV quantile fct}
z_s^p = \mu_s - \frac{1}{\kappa_s\xi_s}\left\{1-[-\log(p)]^{-\xi_s}\right\}.
\end{equation}
That is, $z_s^p$ is the return level associated with the return period $1-1/p$ at site $s$. 

Note that the model formulation in \eqref{eq:gev density} assumes stationarity in time, ignoring e.g. potential effects of climate change. This follows \cite{Wilson&2010} who found no systematic trends over time when analyzing annual maximum flood magnitudes in the Nordic countries. 

\subsubsection{Bayesian hierarchical framework}
The model in \eqref{eq:gev density} assumes a set of site-specific parameters $(\mu_s, \kappa_s, \xi_s)$ at each station $s \in \mc{S}$. The spatial variability is the result of a number of factors related to the variation in terrain and climate which we aim to capture through the covariates $\bs{x}_s$ listed in Table \ref{tab:covariates}. Each of the parameters $\mu_s$,  $\kappa_s$ and $\xi_s$ is specified by a linear model, e.g. for the location parameter
\begin{equation}\label{eq:mu regression}
\mu_s = \bs{x}_s^\top \bs{\theta}^\mu,
\end{equation}
where the first covariate in the vector $\bs{x}_s$ is a constant equal to $1$. Here, we assume that $\bs{\theta}^\mu \in \bs{\Theta}^\mu_{M^\mu}$ for a fixed model $M^\mu$ in which some of the elements of $\bs{\theta}^\mu$ are assumed to take real values, while others may be restricted to be equal to zero. The constraint $\theta^\mu_i = 0$ implies that the $i$th covariate does not influence the location parameter under the model $M^\mu$. In addition to perform inference over the parameter vector $\bs{\theta}^\mu$, we thus also perform inference over the set of possible models $M^\mu$ through Bayesian model averaging as described in \citet[Section 3.3]{Dyrrdal&2015}. We can then assess the posterior marginal inclusion probabilities of each covariate. 

The linear model in \eqref{eq:mu regression} assumes that the variability in the GEV parameters across is fully captured by the variability in the covariates, but in practice there may be additional heterogeneity not directly captured by $\bs{x}_s$. To account for this, we include a site-specific random factor for each of the parameters $\tau_s^\mu$ resulting in the model
\[
\mu_s = \bs{x}_s^\top \bs{\theta}^\mu + \tau_s^\mu,
\]
where the random terms $\tau_s^\mu$ are independent and given by a zero mean Gaussian prior distribution. Our full model can thus be written as 
\begin{align}\label{eq:regional model}
Y_{ts} & \sim GEV(\mu_s,\kappa_s,\xi_s) \nonumber  \\
\mu_s & = \bs{x}_s^\top \bs{\theta}^\mu + \tau_s^\mu  \nonumber \\ 
\eta_s & = \bs{x}_s^\top \bs{\theta}^\kappa + \tau_s^\kappa \\  
\xi_s & = \bs{x}_s^\top \bs{\theta}^\xi + \tau_s^\xi \nonumber \\
\tau_s^\nu & \sim N(0, 1/\alpha^\nu), \quad \nu \in \{ \mu, \kappa, \xi \}, \nonumber 
\end{align}
with $\eta_s = \log(\kappa_s)$. We aim to use uninformative priors, with a gamma distribution for the precision $\alpha^\nu$ and independent standard normal priors for $\bs{\theta}^\nu$ for $\nu \in \{ \mu, \kappa, \xi \}$.

The model in \eqref{eq:regional model} is a slight simplification of the model discussed by \citet{Dyrrdal&2015} as it assumes the random factors $\tau_s^\nu$ for $\nu \in \{ \mu, \kappa, \xi\}$ to be independent across the stations $s$. The inference over the parameters $\left\{\theta^\nu, \alpha^{\nu},\{\tau^{\nu}_s\}_{s \in \mc{S}}\right\}_{\nu \in \{\mu,\kappa,\xi\}}$ can thus be performed as described in \citet{Dyrrdal&2015} using an appropriate modification of the associated package {\tt SpatGEVBMA} \citep{spatgevbma}, available in {\tt R} \citep{Rmanual}. The model in \citet{Dyrrdal&2015} assumes an identity link on the precision parameter $\kappa$. The extension to a logarithmic link requires the calculation of new proposal distributions for the Markov chain Monte Carlo (MCMC) algorithm. This is described in the appendix.  

\subsubsection{Posterior return levels}
We run a Markov chain to return a collection of $R$ samples
\begin{equation}\label{eq:posterior sample}
\left\{\theta^\nu, \alpha^{\nu},\{ \tau^{\nu}_s\}_{s \in \mc{S}}\right\}_{\nu \in \{\mu,\kappa,\xi\}}^{[r]}, \quad r = 1, \dots, R,
\end{equation}
where $R$ is  typically in the range of 50 000 to 100 000, with a suitable number of burn-in samples removed, i.e. the first 10 000 to 20 000 samples. This yields a Markov sample of the parameter set $\{\mu_{s}^{[r]},\kappa_{s}^{[r]},\xi_{s}^{[r]}\}$, including both fixed and random effects. The sample of the corresponding GEV distributions directly yields, by using the GEV quantile function in \eqref{eq:GEV quantile fct}, a sample of quantiles
\[
\{(z_s^p)^{[1]}, \dots, (z_p^s)^{[R]}\}.
\] 
This sample approximates, $pr(z_s^p|\{\bs{y}_s\}_{s \in \mc{S}})$, the posterior distribution of the $p$th return level at the site $s \in \mc{S}$.  Here, $\bs{y}_s$ denotes the vector of maximum floods at site $s$ for all years $t$ for which measurements are available. Given this sample, it is straightforward to derive approximations for the posterior mean and median with point-wise posterior credible intervals for the $p$th return level.   

The Markov sample also constitutes a mixture distribution based on the GEV density function in \eqref{eq:gev density}
\begin{equation}
pr(y_{s}|\{\bs{y}_s\}_{s \in \mc{S}}) = \frac{1}{R} \sum ^{R}_{r=1} pr_{GEV}(y_{s}|\mu_{s}^{[r]},\kappa_{s}^{[r]},\xi_{s}^{[r]}),
\label{eq:mixture}
\end{equation}
approximating the posterior predictive distribution of a future observation $y_{st}$. Such mixture distributions do not have an explicit quantile function and the return level is found by simulating a number of observations from each mixture component, using the empirical quantile of observations pooled over all mixture components as an approximation. Due to the large number of Markov samples, only around 10 to 100 simulated observations from each component are needed to achieve a good approximation of the true return level. The Equation \eqref{eq:mixture} implies BMA, given that the predictive distribution is averaged of different models. 

Now assume we are interested in estimating the $p$th return level at a new site $s_0 \notin \mc{S}$ not used to estimate the model, but for which the covariates $\bs{x}_{s_0}$ listed in Table \ref{tab:covariates} are available. The $p$th return level of the posterior predictive distribution for site $s_{0}$ is given by the empirical $p$th quantile found when combining simulated observations from all mixture components, based on the fixed effects $\bs{x}_{s_0}^\top (\bs{\theta}^\mu)^{[r]}$, $\kappa_{s_0}^{[r]}$ and $\xi_{s_0}^{[r]}$. The quantile of the predictive distribution will be close, but not identical, to (the approximation of) the median of the posterior distribution of the $p$th return level. 

The uncertainty is quantified by the point-wise 80 \% posterior credible intervals of the quantiles corresponding to a posterior sample of the three GEV parameters, given as follows
\begin{enumerate}
\item[(i)] sample $(\tau_{s_0}^\mu)^{[r]} \sim N(0, 1/(\alpha^\mu)^{[r]})$;
\item[(ii)] set $\mu^{[r]}_{s_0} = \bs{x}_{s_0}^\top (\bs{\theta}^\mu)^{[r]} + (\tau_s^\mu)^{[r]}$,
\end{enumerate}
and similar for $\kappa_{s_0}^{[r]}$ and $\xi_{s_0}^{[r]}$. The additional sampling of random effects gives a better (out-of-sample) calibration. Note that in general there is a higher level of uncertainty for a new site $s_{0}$, than a site $s \in \mc{S}$ used to estimate the model, due to the independent sampling step in (i). This additional sampling of the random effects ensures that the uncertainty is not underestimated.

\subsection{Local GEV model}
\begin{figure}[!hbpt]
\centering
\includegraphics[width=\textwidth]{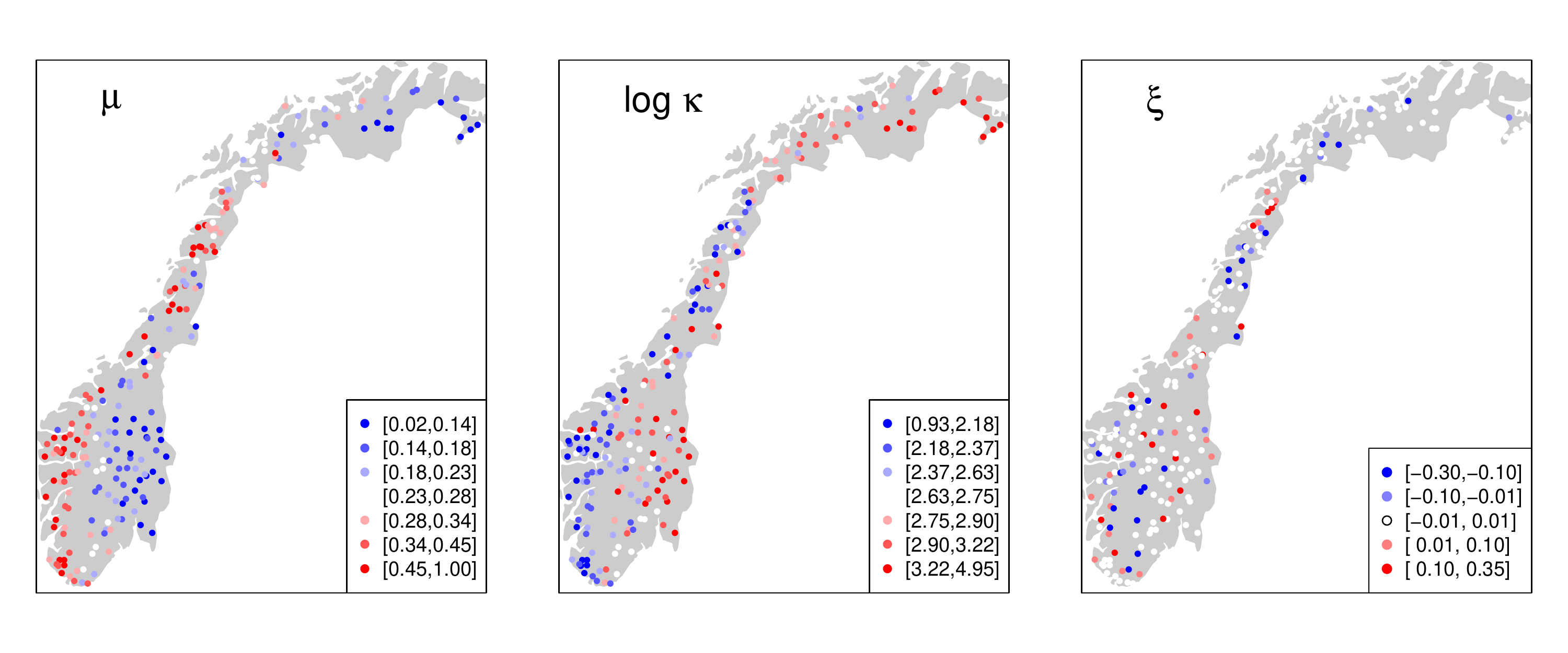}
\caption{Spatial distribution of the median parameter estimates from the local GEV model.}\label{fig:MAPlocalGEV}
\end{figure}

We compare the regional model to a local non-hierarchical GEV model for on-site analysis without any spatial structure. In this setting, we assume that the annual maxima are described by Equation \eqref{eq:gev density} with independent parameters, $\mu_s,\kappa_s$, and $\xi_s$, not sharing informative covariates. The unknown parameters are estimated by MCMC within a Bayesian framework such that $\mu_s,\kappa_s$, and $\xi_s$ is updated in turn, see \citet{Steinbakk&2016} for details on the prior distributions and their updates. For stations with data series less than 50 years, the shape parameter $\xi_s$ is assumed to be zero and a Gumbel distribution is fitted instead of the GEV distribution according to FFA recommendations for Norway \citep{Midttomme&2011, Castellarin&2012}. Figure \ref{fig:MAPlocalGEV} shows the spatial distribution of the estimated parameters from the local GEV model. The map of the shape parameter $\xi_s$ reveals that a considerable proportion of sites are estimated using a Gumbel distribution due to short data series.

\section{Current regional model for Norway}
The Norwegian Water Resources and Energy Directorate (NVE) is responsible for determining guidelines for flood frequency analysis in Norway. The current regional framework for estimating flood return levels were established by \citet{Saelthun&1997}, see also \cite{Castellarin&2012}. The return values are found by first estimating the index flood based on geographical and hydrological parameters, and second, using a growth curve to scale up the index flood to a certain design flood. The two steps are determined by different procedures.  Both the model for index flood and the growth curve are based on initially dividing sites into distinct flood types (spring, autumn, glacier or all year floods) with geographical regions; four different zones for spring floods and three zones of autumn floods. Each subdivision then has a separate regression model and relevant covariates. An overview of the range of covariates, with descriptions, used in the separate regression models is given in Table \ref{tab:oldModel}. 

\begin{table}[!hbpt]
\centering
\caption{Overview of key parameters for computing index flood in the current system of NVE}\label{tab:oldModel}
\begin{tabular}{l c } 
\hline
Parameter-name &  Unit  \\
\hline
Catchment area & km$^2$ \\
Mean specific annual runoff  & $l/s$ per $km^{ 2}$ \\
Mean annual precipitation  & mm \\
Effective lake  & \% \\
Exposed bedrock  & \% \\
Catchment length  &km \\
Gradient of the main river & m/km \\
\hline
\end{tabular}
\end{table}

One difficulty with the current regional model is the lack of a rigorous definition of the flood regions. It can be difficult to determine the appropriate region of a new site, and the estimated return values vary with the choice of region. Hence it will be highly beneficial to model the regional or spatial characteristics of floods as a continuum in a more complex regression model. In addition, the growth curve is only available for certain predetermined return periods. 

\section{Model validation}\label{sec:validation}
To validate the models we follow \cite{Renard&2013} assessing reliability, or calibration, and stability. Reliability describes the consistency between validation data (data not used for calibration) and FFA predictions. A reliable, or well-calibrated, model should yield an estimated distribution close to the unknown true distribution of the data. Stability, on the other hand, quantifies
the ability of the model to yield similar estimates when calibration data change. 

We assess the predictive power of the regional model through a cross-validation study, such that reliability is assessed through the consistency between predictions and hold-out data. Due to the heavy computational burden of the Markov sampling in the hierarchical model, a smaller number of sites, specifically 27 stations, were selected by experts to represent the range of different sites in Norway. We employ a leave-one-out cross-validation scheme, where all data for each of the 27 stations, in turn, are left out of the model fitting. The distribution of the random effects gives the main difference between in-sample and out-of-sample predictions, as the in-sample parameter estimates allow for, and usually have, correlated random effects, while the random effects are independently drawn for the out-of-sample estimates. 

\subsection{Reliability}
The main reliability assessment tool is the probability integral transforms (PITs), displayed graphically by histograms and probability-probability (PP) plots. If observations follow the estimated distribution, the PIT will be uniformly distributed \citep{Dawid1984}
$$ \hat F(y_{st}) \sim U([0,1]).$$ 
Histograms of PIT values can be assessed at a local level if the data series are long, whereas histograms with too few observations do not allow for a useful graphical assessment. Instead one should assess regional average reliability by combining PIT values from several or all locations in a single histogram. The reliability of individual stations is assessed by PP plots displaying the observed empirical distribution of PIT values against the theoretical uniform distribution. The values should to the largest degree follow a one-to-one line, and deviation will indicate bad reliability such as over- or underestimation compared to observed values.

We assess the reliability in the tail of the predictive distribution using proper scoring rules, in particularly the quantile score, see e.g. \citet{FriederichsHense2007}, \citet{GneitingRaftery2007} and references therein. If denoting the predictive distribution by $F$ and the realized observation by $y$, the quantile score is given by
\[
s_Q(F,y|\tau) = (y - F^{-1}(\tau))(\tau - \mathbbm{1}\{ y \leq F^{-1}(\tau)\}),
\]
for a specific probability level $\tau \in (0,1)$. Alternatively, one could use the weighted continuous ranked probability score integrating over all quantiles greater than some threshold, say the 50-year return level \citep{GneitingRanjan2011}. But as the current regional model used in Norway is only (easily) available for certain pre-determined return periods, models are compared using the quantile scores.

While histograms and PP plots of PIT values only assess the reliability of the predictions, scoring rules such as the quantile score simultaneously assess several aspects of the predictive distributions, see e.g. \cite{Stephenson&2008} and \cite{BentzienFriederichs2014}.  That is, by conditioning (stratifying) on the predicted probabilities, the scores may be decomposed into the sum of three components: reliability, resolution and uncertainty. The resolution is related to the information content of the prediction model. It describes the variability of the observations under different forecasts and indicates whether the prediction model can discriminate between different outcomes of an observation. In our setting, a prediction model with a good resolution is able to discriminate between locations with different flood characteristics while a prediction model with no resolution would issue the same predictive distribution at all locations. The uncertainty component refers to the variability in the observations and is thus identical for competing models when assessed under the same data set.

\subsection{Stability}
Within a comparison framework, \cite{Renard&2013} advised to first assess reliability, and then use stability to further discriminate between models if several models are equally reliable. The stability quantifies to which degree the statistical model yields similar predictive distributions when trained on different, but identically distributed, data sets. This is a property solely of the statistical model, thus arbitrarily large return periods can be assessed. A general procedure is as follows: First, the statistical model is fitted using all available data, yielding an estimated predictive distribution $\hat{G}$, followed by estimation on a set of leave-one-out scenarios, yielding estimated predictive distributions $\hat{F}_1, \ldots \hat{F}_n$. Then each $\hat{F}_i$ is compared to $\hat{G}$ using some measure of divergence, giving an average or maximum divergence for $i = 1, \ldots, n$. We will assess the stability of our model through the variability of parameter estimates, in particular the intercept term, over the leave-one-out cross-validation scheme.

\section{Results}
This section shows the results from the aforementioned models to predict annual flood maxima in Norway. We first assess the reliability of the regional model following the cross-validation study and then present in-depth results for certain stations, selected to showcase the range of individual site behavior.    

\subsection{Bayesian regional model}
Relevant covariates for the regional model were first explored by assessing relationships between covariates and the estimated $\mu_s$, $\kappa_s$ and $\xi_s$ from the local model. This revealed that some covariates needed to be transformed, or be combined into variables. To decide on the specific set of covariates, we used stepwise linear regression optimizing AIC \citep{siotani1985modern}, with the index (mean level) flood for each station as the response. The resulting best model contained 13 covariates, seen in Table \ref{tab:covariates}, and the selected variables overlap to a large degree with the covariates used by the current regional mode, for details see Table \ref{tab:oldModel}. New covariates not considered in the current framework are the average fraction of rain versus snowmelt, the meteorological variables (i.e. rain in April and August and snowmelt in March) and, in particular, longitude and latitude. The regional model was run using 100,000 MCMC iterations and 20,000 burn-in samples, with the posterior marginal inclusion probability of each covariate given in Table \ref{tab:inclProbRegModel}. 

\begin{table}[!hbpt]
\centering
\caption{Inclusion probability (\%) for the covariates for models for location parameter $\mu_s$, scale parameter $\kappa_s$ and
shape parameter $\xi_s$.\label{tab:inclProbRegModel}}
\begin{tabular}{l r r r}
\hline
&\multicolumn{1}{c}{$\mu_s$}&\multicolumn{1}{c}{$\kappa_s$}&\multicolumn{1}{c}{$\xi_s$}\\
\hline
Constant & 100 & 100 & 100 \\ 
  Longitude & 53 & 99 & 6 \\ 
  Latitude & 84 & 100 & 8 \\ 
  Percent of effective lake & 98 & 100 & 2 \\ 
  Inflow & 6 & 11 & 9 \\ 
  Average fraction of rain & 3 & 22 & 58 \\ 
  Catchment area & 2 & 5 & 12 \\ 
  Average rain in April & 42 & 75 & 5 \\ 
  Average rain in August & 100 & 100 & 5 \\ 
  Snow melting in March & 8 & 16 & 4 \\ 
  Catchment gradient & 45 & 12 & 2 \\ 
  Percent of bedrock & 22 & 8 & 2 \\ 
  Relative area to length & 13 & 95 & 11 \\ 
   \hline
\hline
\end{tabular}
\end{table}

\begin{figure}[!hbpt]
\includegraphics[width=\columnwidth]{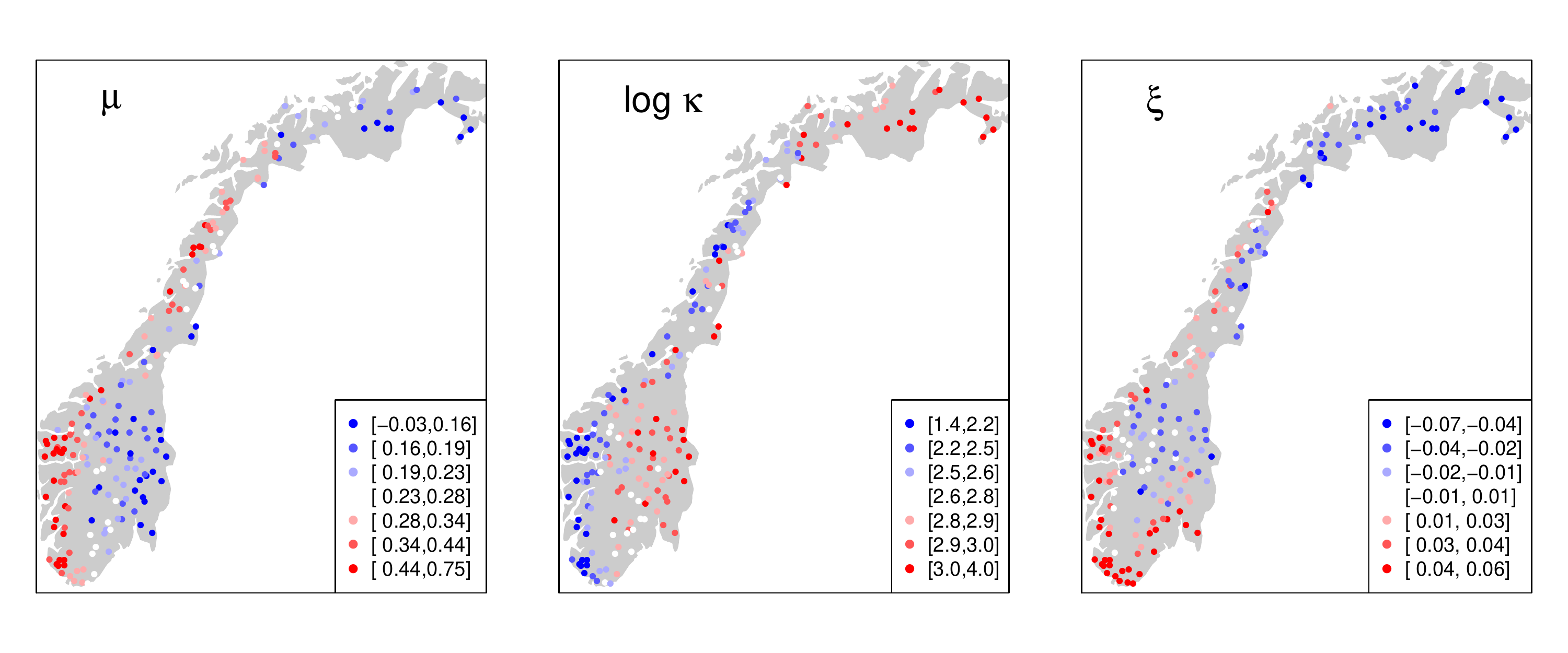}
\caption{Spatial distribution of the median fixed effects per site, for 203 stations, for the three GEV parameters, $\mu_{s}^{[r]},\kappa_{s}^{[r]}$ and $\xi_{s}^{[r]}$.}%
\label{fig:fixedEffects}%
\end{figure}

It is seen that spatial location of a station, in terms of longitude and latitude, is crucial for all GEV parameters, thereby reflecting the importance of flood regions and the spatial similarity between flood sites. Maps of the median estimate of the fixed effects for $\mu_s$, $\kappa_s$ and $\xi_s$ are shown in Figure \ref{fig:fixedEffects}. There are clear spatial structures with an east to west and north to south gradient in both the location and the scaling parameter, while the shape parameter mainly displays a difference between coast and inland. From Figure \ref{fig:MAPlocalGEV} one can see that the estimated fixed effects for $\mu_S$ and $\kappa_S$ agree with the distribution of the parameters from the local model, while the pattern of $\xi_S$ is not observed in the local model. These results highlight that pooling of information across catchment sites via the linear regression model is highly beneficial. 

Further, both the parameters  $\mu_s$ and $\kappa_s$ are explained by the percentage of effective lake, as a dominant lake in the catchment area can dampen both the mean flood level and the yearly variations, and by the average rain in April and August, as more rain will increase both average flood levels and variability. The derived covariate, catchment area divided by the basin length, is seen to significantly affect the scaling $\kappa_s$, suggesting that ``long'' catchments (elongated along the length) experience less variability possibly due to a dampening effect. The percent of bedrock and the catchment gradient, on the other hand, influence mainly the location parameter $\mu_s$, as a larger degree of mountainous terrain within a catchment and a steeper gradient will increase the average flood level. Lastly, the shape parameter $\xi_s$ is mainly explained by the average fraction of rain in the flood, and to a smaller degree the catchment area, the relative area to length and the inflow. In areas with a smaller fraction of rain compared to snowmelt, the annual maximum flood is more often a spring flood caused by snowmelt, which will to a larger degree be limited by an upper threshold. This characteristic would be accounted for by a negative shape parameter. 
 
\begin{figure}[!hbpt]
\includegraphics[width=\columnwidth]{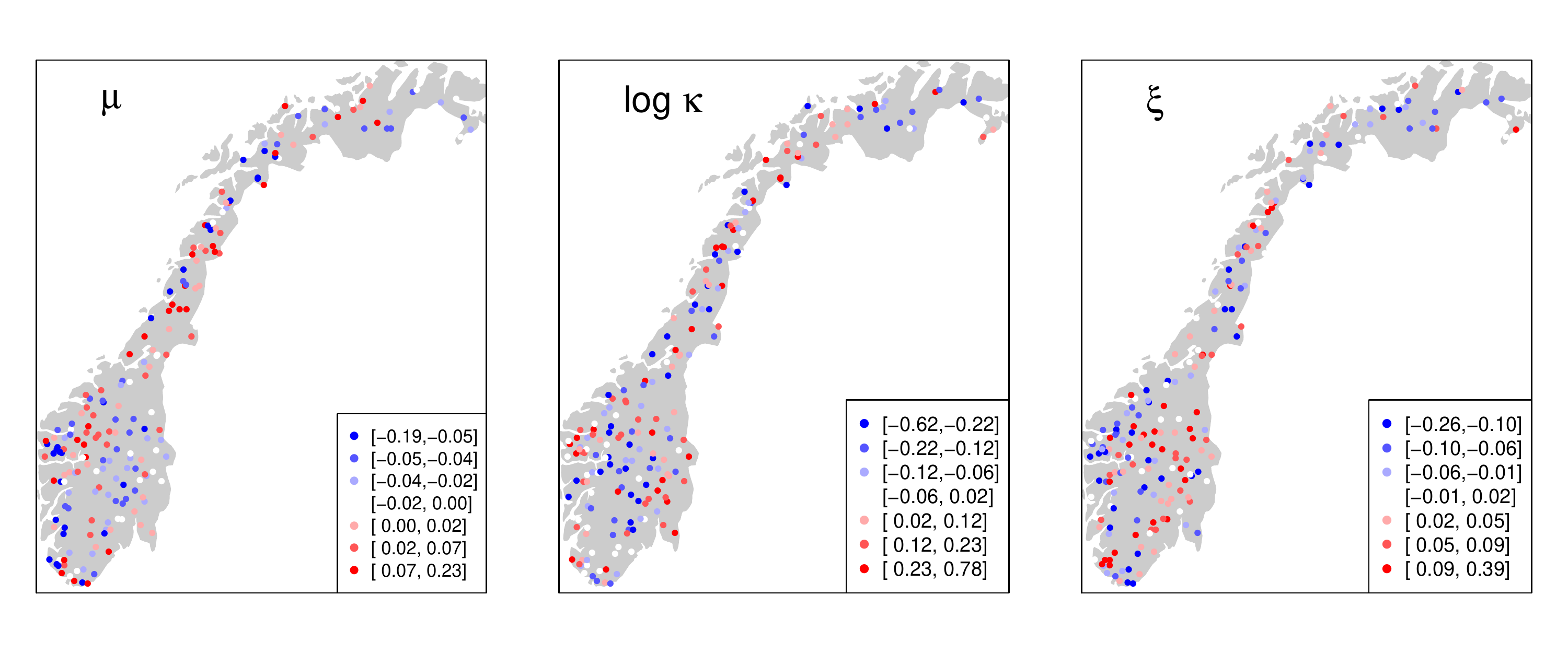}%
\caption{Spatial distribution of the median random effects per site, for 203 stations, for the three GEV parameters, $\mu_{s}^{[r]},\kappa_{s}^{[r]}$ and $\xi_{s}^{[r]}$.}%
\label{fig:randomEffects}%
\end{figure}

Figure \ref{fig:randomEffects} shows the median estimates of random effects for the three parameters, per station, and could reveal whether additional spatial effects need to be accounted for. Overall, the random effects seem to be spatially independent, apart from scattered, but small, clusters present in all three parameters which could be due to  river networks or regional catchment areas.  

\begin{figure}[!hbpt]
\includegraphics[width=\columnwidth]{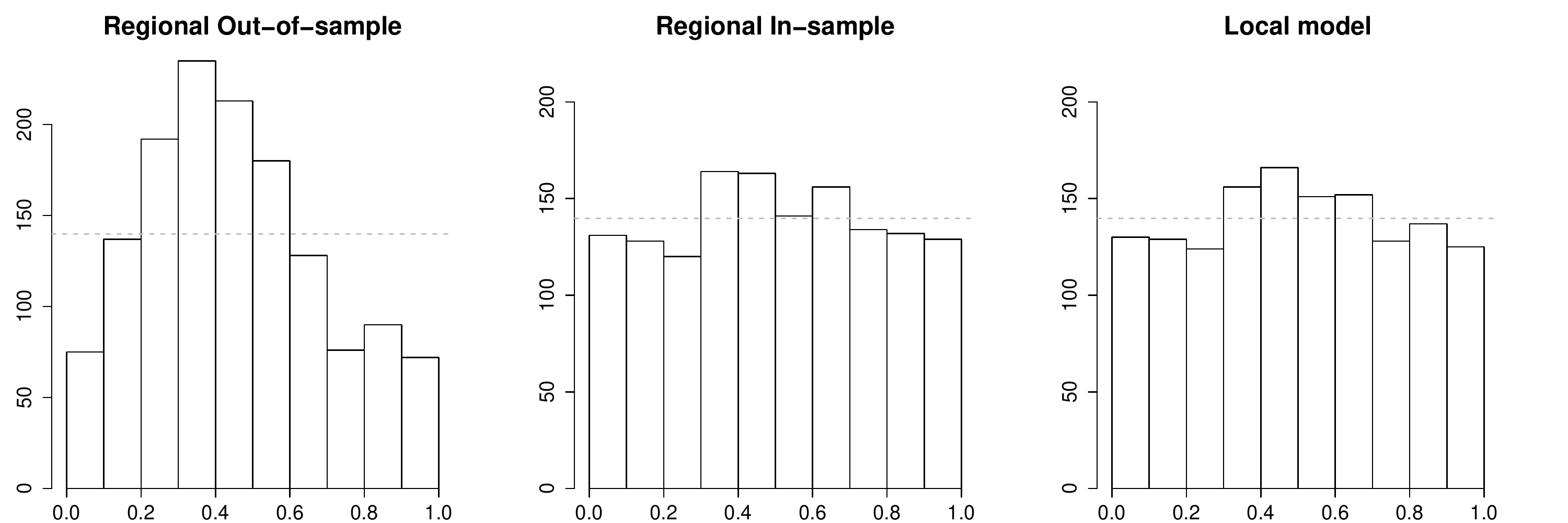}%
\caption{Histograms of PIT values for all observations from the 27 cross-validation stations, a total of 1305 observations. The regional out-of-sample model shows an excess of PIT values away to 0 and 1.}%
\label{fig:histogramPIT}%
\end{figure}

\begin{figure}[!hbpt]
\includegraphics[width=\columnwidth]{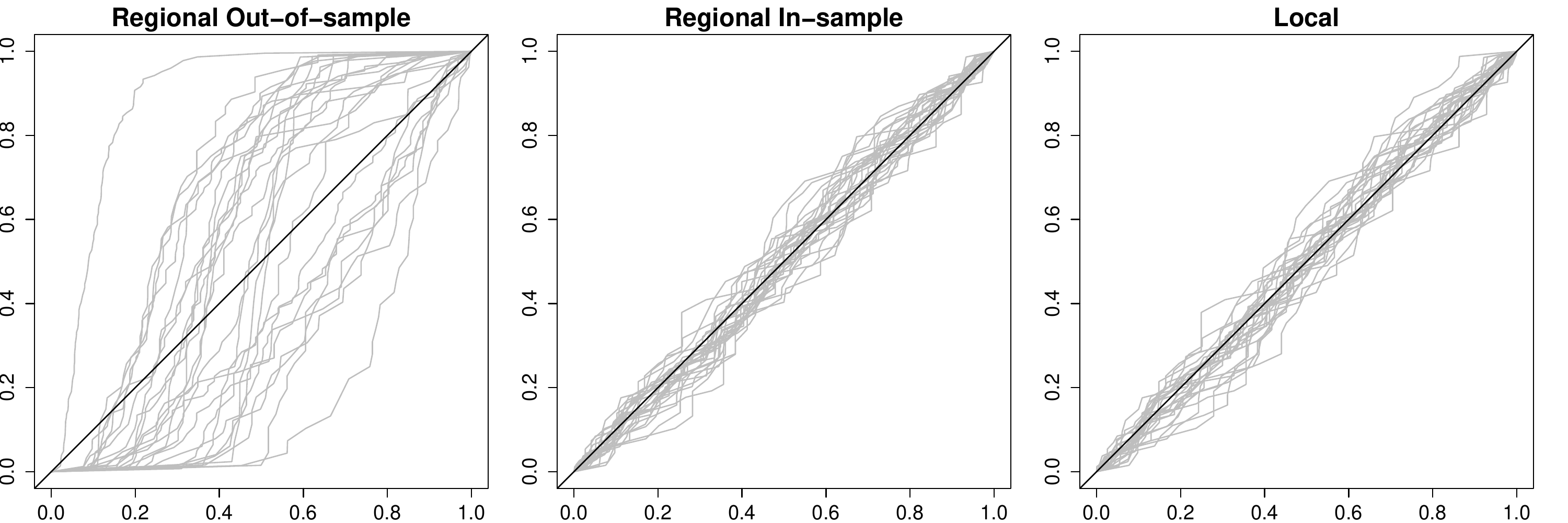}%
\caption{Probability-probability plots of the PIT values for the 27 cross-validation stations. For the regional out-of-sample model some stations are highly overestimated, while others are underestimated.}%
\label{fig:ppvaluesPIT}%
\end{figure}

\begin{figure}[!hbpt]
\includegraphics[width=\columnwidth]{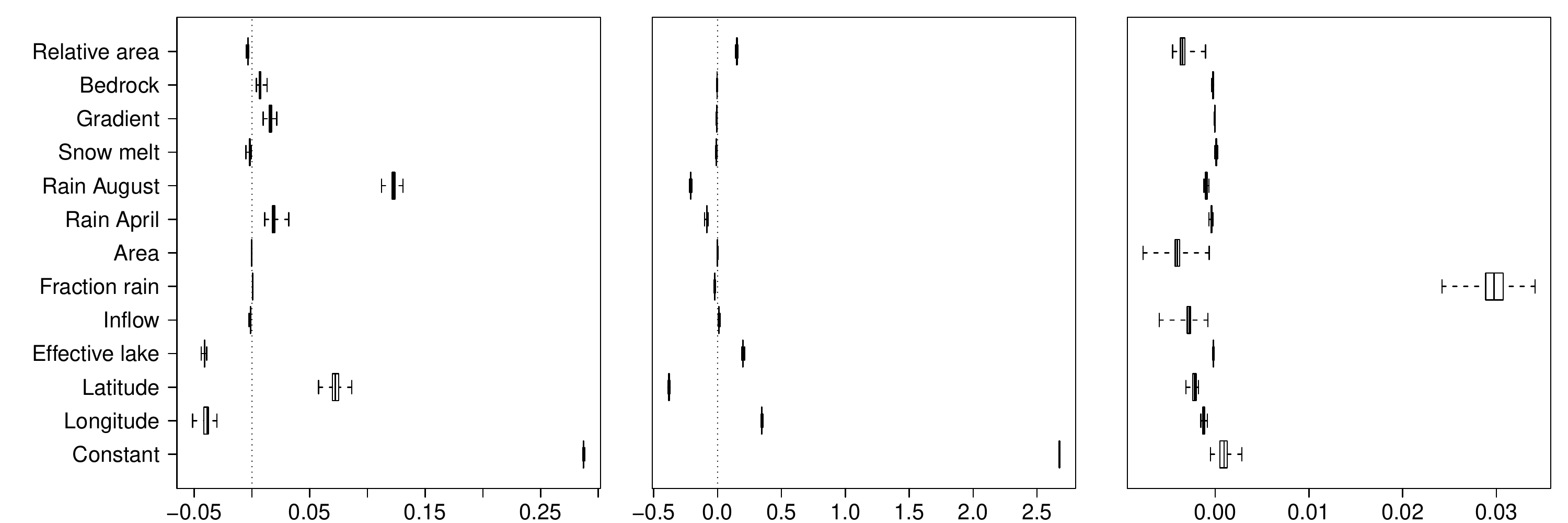}%
\caption{Assessment of stability: Box plots displaying the variability of the posterior mean regression coefficients over the 27 cross-validation models, shown for each of the 13 covariates behind $\mu,\log \kappa,$ and $\xi$.}%
\label{fig:stability}%
\end{figure}

We validate the model following Section \ref{sec:validation} and start by comparing histograms of aggregated PIT values for the out-of-sample and in-sample regional models, and the local model. Figure \ref{fig:histogramPIT} shows that the local and in-sample regional models are well calibrated over the 27 selected stations, while the out-of-sample model is somewhat overdispersive. The PP plots in Figure \ref{fig:ppvaluesPIT} show that the regional model will severely over- and underestimate the return levels of some out-of-sample stations, while the predictive distribution of most stations is seen to be well calibrated. We assess the stability of the model through the variability of the fixed effects in each of the three parameters. Figure \ref{fig:stability} shows box plots of the mean regression coefficients over the 27 leave-one-out cross-validation models, for each of the 13 selected covariates and the three parameters. All estimates are seen to be very stable, in particular for $\mu$ and $\kappa$, and only the coefficient estimates of the rain contribution and area for $\xi$ are slightly less stable.

\begin{figure}[!hbpt]
\centering
\includegraphics[width=0.5\columnwidth]{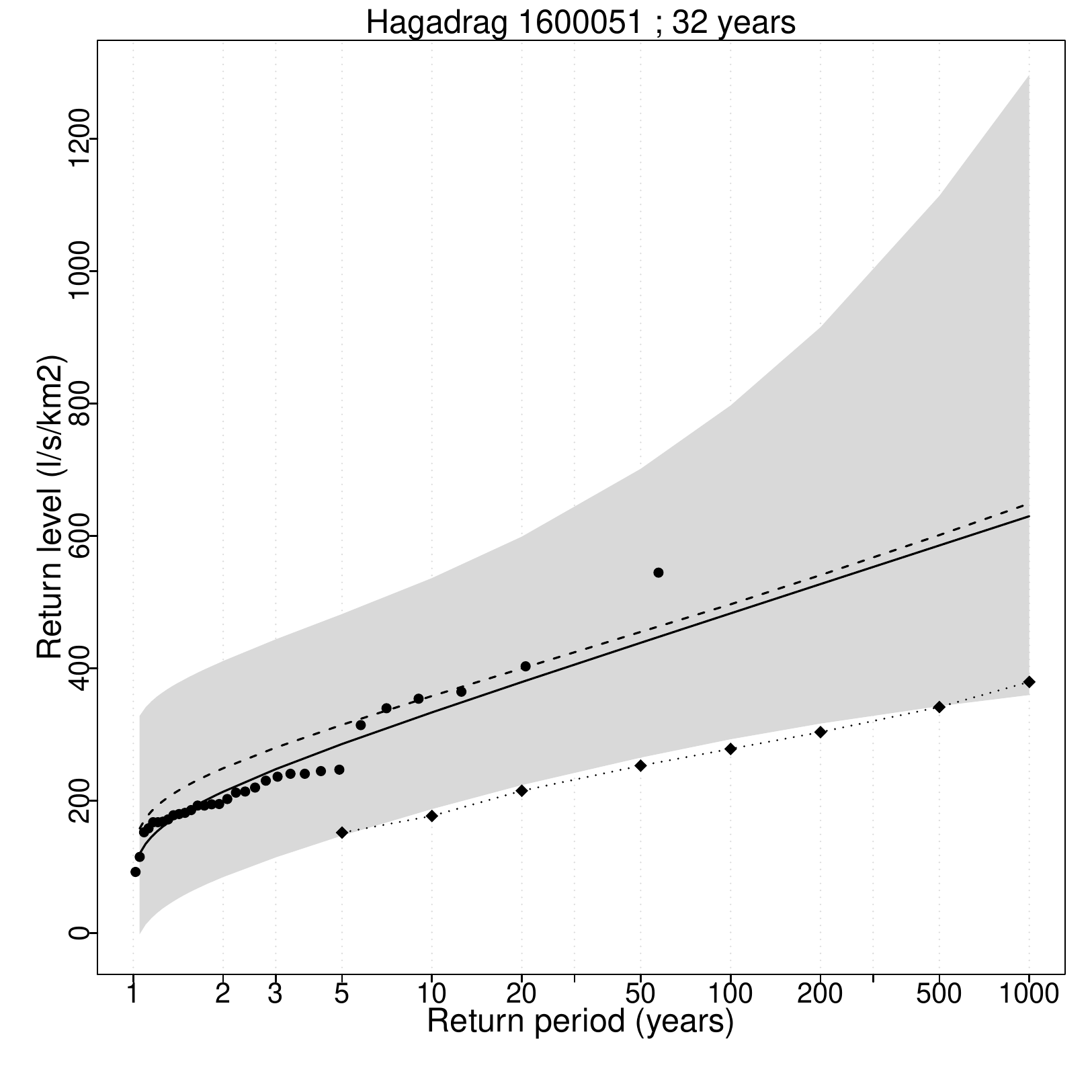}\\
\includegraphics[width=0.48\columnwidth]{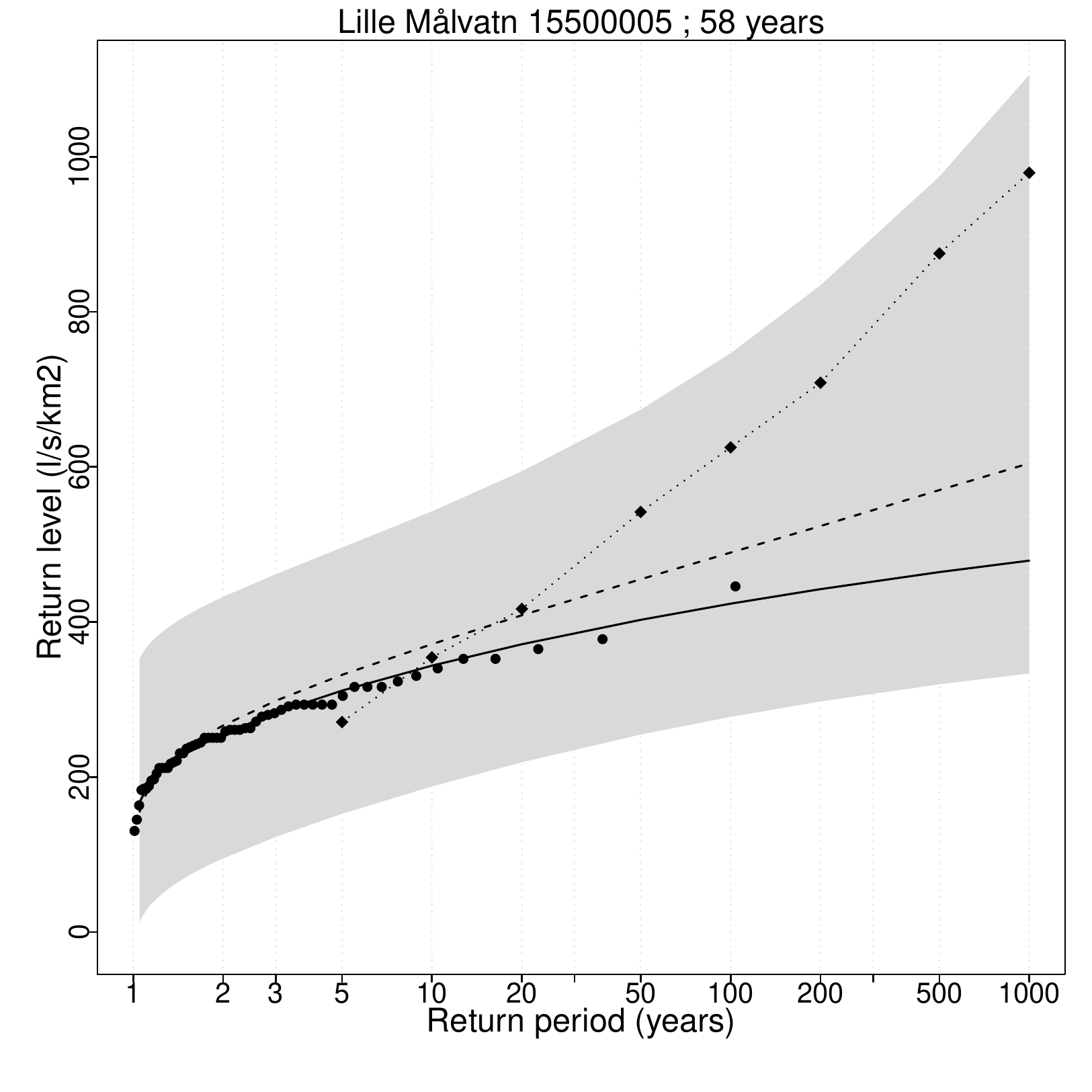}
\includegraphics[width=0.48\columnwidth]{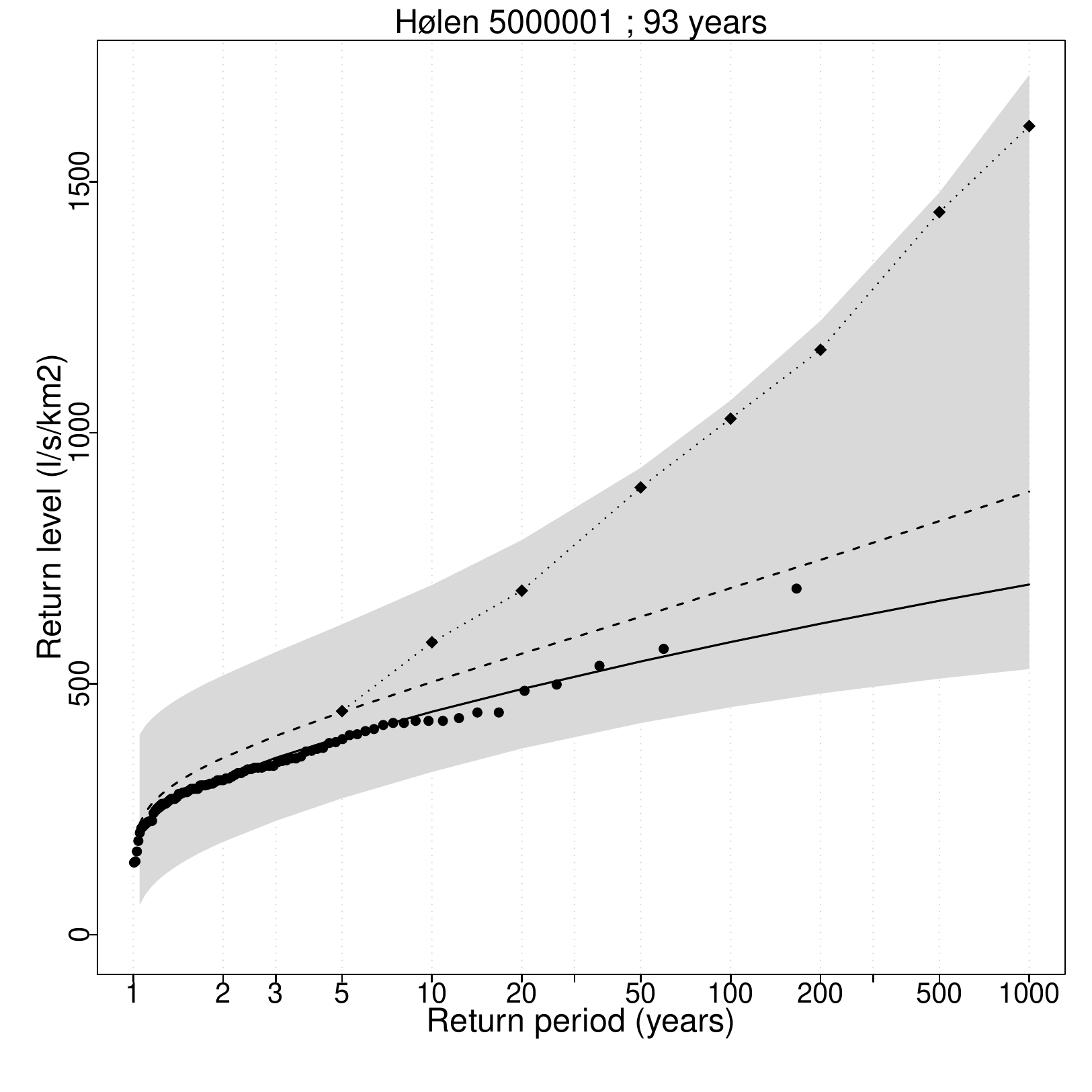}
\caption{Estimated return levels for three stations (32, 56 and 93 years of observations, respectively) showing good agreement between the regional and local model. Black lines: Local Bayesian model. Dashed lines: Posterior predictive distribution. Gray area: 80\% credibility interval for the posterior quantile distribution. Dotted line with squares: Standard regional model. Black dots: Data.}%
\label{fig:goodFit}%
\end{figure}

To compare the fit of the different models with the observed data, we explore the results of six selected sites. Figure \ref{fig:goodFit} shows the (out-of-sample) estimated return level plotted against the return period (on a log scale), for three sites with short ($n = 32$), medium ($n = 56$) and long ($n =93$) data series. The figure shows the observed data (black dots), the median estimate of the local model (black lines), predictive distribution (dashed lines) and pointwise 80\% credible bands (gray area) of the return level for the regional model and the current regional model (in dotted lines with squares). The displayed stations have been selected to reflect a good fit by the regional model to the observed data, such that the regional predictive distribution gives a 1000 year return level in agreement with the local model. It is seen that for these three selected sites the new regional model greatly improves on the regional model currently used in Norway. In addition, it is also clear that the current regional model stays within the 80 \% credible bands of the new regional model.   

\begin{figure}[!hbpt]
\centering
\includegraphics[width=0.5\columnwidth]{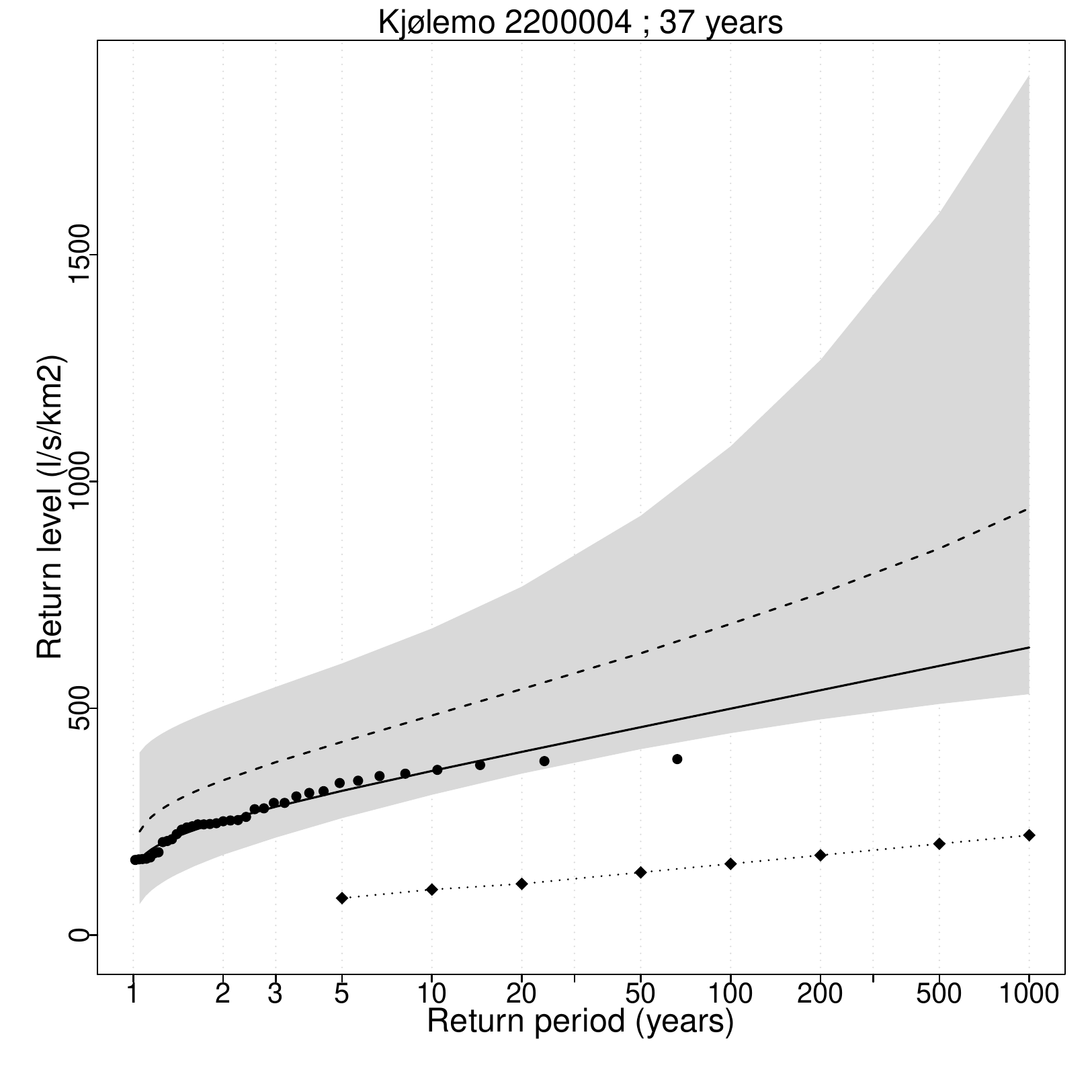}\\
\includegraphics[width=0.48\columnwidth]{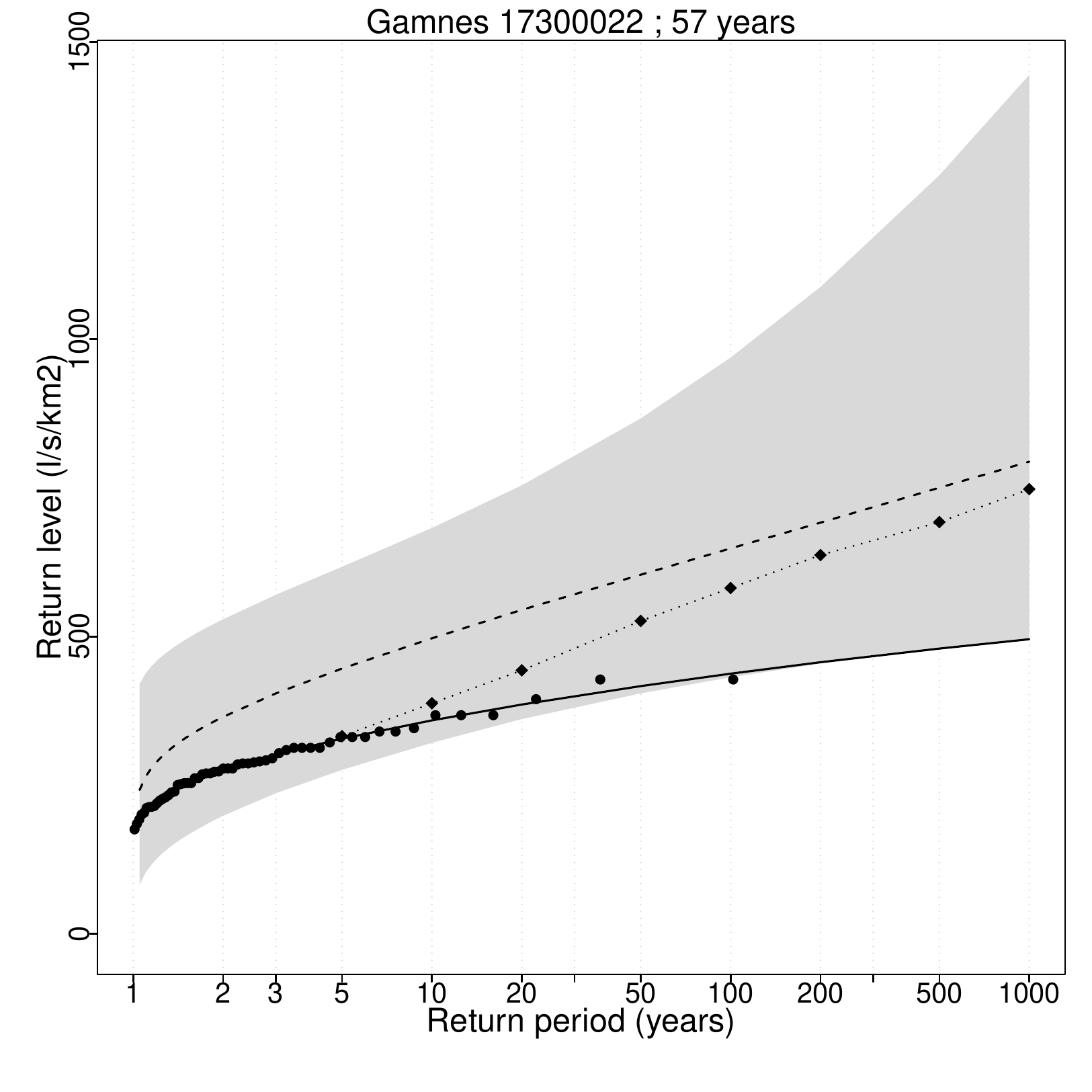}
\includegraphics[width=0.48\columnwidth]{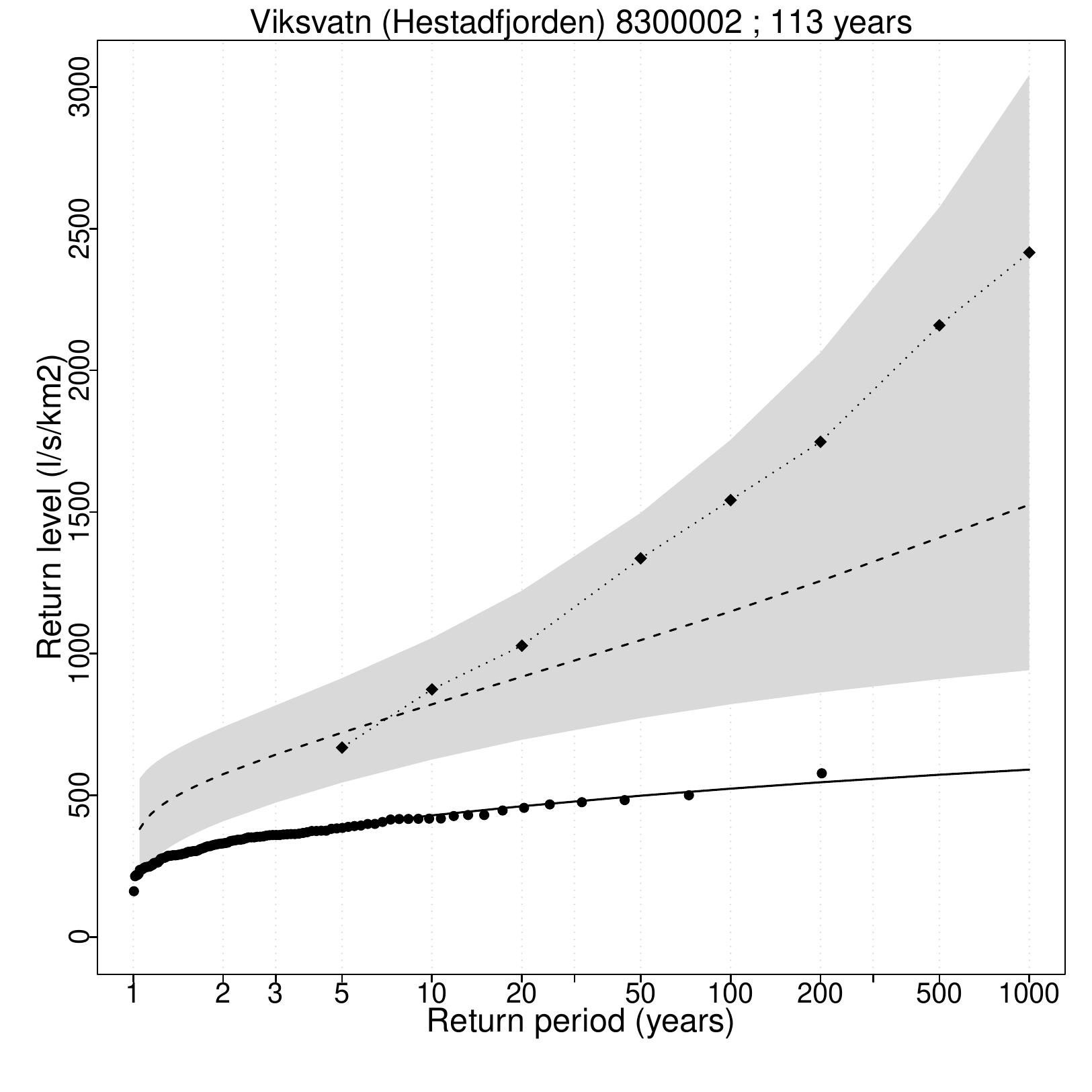} 
\caption{Estimated return levels for three stations (32, 56 and 93 years of observations, respectively) with a bad agreement between the regional and local model. Black lines: Local Bayesian model. Dashed lines: Posterior predictive distribution. Gray area: 80\% credibility interval for the posterior quantile distribution. Dotted line with squares: Standard regional model. Black dots: Data.}%
\label{fig:badFit}%
\end{figure}

Figure \ref{fig:badFit} shows the same out-of-sample return value plots for three sites with a short ($n=37$), medium ($n=57$) and long ($n=113$) data series, but selected to reflect a bad agreement between the regional model and observed data. In all three sites the regional model overestimates the return levels compared to the local model, but in all three cases our regional model gives a better fit than the current regional model. Table \ref{tab:quantileScore} displays the average quantile scores for the out-of-sample predictions for the 27 selected stations, comparing the regional model to the local model and the regional model currently used by NVE. For all return periods, $T= 10,50,100$, the new regional model performs significantly better than the current regional model, but not as well as the local model. 

\begin{table}[!hbpt]
\centering
\caption{Quantile scores for return periods $T=10$, 50 and 100 comparing the regional model with the local and the regional model currently in use at NVE \citep{Saelthun&1997}. The 90\% uncertainty intervals are obtained by bootstrapping.}\
\label{tab:quantileScore}
\begin{tabular}{rllllll}
  \hline
& &$T=10$ &  & $T=50$ & & $T=100$   \\ 
  \hline
NVE model & 6.71 & [6.23, \, 7.20] & 4.04 & [3.63, \, 4.46] & 3.17 & [2.80, \, 3.55] \\ 
  Regional model & 2.94 & [2.72, \, 3.17] & 1.01 & [0.87, \, 1.17] & 0.62 & [0.51, \, 0.75] \\ 
  Local model & 1.95 & [1.80, \, 2.11] & 0.63 & [0.57, \, 0.70] & 0.36 & [0.32, \, 0.40] \\ 
   \hline
\end{tabular}
\end{table}

\section{Discussion}
We have, in accordance with our main objective, developed a regional model for extreme flood estimation to be used when little or no data are available at a catchment site. The model is a Bayesian hierarchical framework with site-specific GEV parameters based on geographical, hydrological and meteorological covariates, such that information is shared across catchment sites. We have identified 12 important predictors of this type to describe the local variation in the model parameters. By evaluating quantile scores for different return periods, it is seen that the Bayesian regional model gives better predictive performance than the current regional model used by NVE in Norway. Recent work by \cite{Yan&Moradkhani2016} also supports that methods utilizing Bayesian hierarchical models and model averaging are beneficial for analyzing extreme flood data when emphasizing the quantification of extreme flood uncertainty.

While the new regional model significantly improves upon the current NVE model, a reliability analysis at 27 out-of-sample stations reveals that the model is, overall, somewhat overdispersive.  A further analysis of the performance at individual stations indicates that the selected set of covariates may not be able to pick up the location-specific effects for all locations. Among the selected stations, the regional model tends to more often provide a higher estimates of flood sizes for high return periods than the local model. The Viksvatn station by the Gaular river in Sogn is one example of a catchment site where the regional model highly overestimates the flood distribution compared to the local model which is based on 113 years of observed annual floods, as seen in Figure \ref{fig:badFit}. However, in other unrelated analyzes, we have found that it appears that recorded level of rain at catchment sites in the Sogn area are consistently estimated too high. As the average rain in August is one of the most influential predictors of the overall flood level, any overestimation of this covariate will have an extreme impact on predicted return levels. This highlights the importance of good data quality for achieving reliable return level predictions.

The covariate describing the average contribution of rain and snowmelt in the floods is found to be the best predictor of the shape parameter $\xi_s$. However, the exact calculation of this covariate requires some observed flood data. While the covariate may be estimated solely based on available meteorological data products such as the SeNorge data product, it remains to be assessed how much additional uncertainty is introduced. 

\section*{Acknowledgments}
This work was jointly funded by The Research Council of Norway and Energy Norway through grant 235710 (FlomQ). The flood and hydrological data were extracted from the national hydrological database at the Norwegian Water Resources and Energy Directorate and are listed in the supplementary materials.  Climatological data, derived from the data product SeNorge, are available at \url{www.senorge.com}.The source code for estimating the flood frequency distributions is implemented in the statistical programming language R (\url{http://www.R-project.org}) as a part of the R package \verb|SpatGEVBMA| which is available on GitHub at \url{http://github.com/NorskRegnesentral/SpatGEVBMA}.

\bibliographystyle{chicago}
\bibliography{flom}

\clearpage

\appendix

\section{Hierarchical model with a $\log$ link on the precision}

This section discusses an extension to the MCMC algorithm described in \citet{Dyrrdal&2015} where the regression equation for the precision parameter is defined with $\log \kappa$ as the response variable. In general, assume we want to update a parameter $\nu$ in our model, where $\nu$ is the current value. We draw a new value $\nu'$ from a proposal distribution $pr(\nu' | \nu, \cdot)$ and accept the proposal with probability $\min \{ r, 1 \}$ where 
\[
r = \frac{pr(\bs{y}| \nu', \cdot) pr(\nu' | \cdot) pr(\nu | \nu', \cdot)}{pr(\bs{y} | \nu, \cdot) pr (\nu| \cdot) pr(\nu' | \nu, \cdot)}.
\]
Here, $pr(\bs{y} | \nu, \cdot)$ denotes the likelihood of the full data set $\bs{y}$ which depends on $\nu$ and potentially other parameters which are kept fixed throughout, and $pr( \nu | \cdot)$ is the prior distribution of $\nu$ which similarly might depend on the other parts of the model.  Given the complexity of the model, it is vital to design efficient proposal distributions which return good proposals and are robust in that they do not require fine-tuning for each individual data set.  

For designing the proposal distribution, we employ a Gaussian approximation \citep[][Ch. 4.4]{RueHeld2005}.  Assume that the posterior distribution of the parameter $\nu'$ is written on the form
\[
pr(\nu'| \cdot) \propto \exp \big( f(\nu') \big),
\]
for some function $f$.  A quadratic Taylor expansion of the log-posterior $f(\nu')$ around the value $\nu$ gives 
\begin{align*}
f(\nu') & \approx f(\nu) + f'(\nu) ( \nu' - \nu) + \frac{1}{2} f''(\nu) (\nu' - \nu)^2 \\
& = a + b \nu' - \frac{1}{2} c (\nu')^2, 
\end{align*}
where $b = f'(\nu) - f''(\nu) \nu$ and $c = -f''(\nu)$.  The posterior distribution $pr(\nu' | \cdot)$ may now be approximated by 
\[
\widetilde{pr}(\nu' | \cdot) \propto \exp \Big( -\frac{1}{2} c (\nu')^2 + b \nu' \Big), 
\]
the density of the Gaussian distribution $\mc{N}(b/c, c^{-1})$. We thus choose $\mc{N}(b/c, c^{-1})$ as our proposal distribution. This implies that in order to update the model in \citet{Dyrrdal&2015} to include a logarithmic link for the precision $\kappa$, we need to calculate the first two derivatives of the log likelihood function $\log pr(y_{ts}|\tau_s^{\kappa}, \cdot)$ with respect to the random effect $\tau_s^{\kappa}$. 

\subsection{The case $\xi \neq 0$}

We have $\kappa_s = \exp(\eta_s)$, where $\eta_s = \bs{x}_s^\top \bs{\theta}^\kappa + \tau_s^\kappa$ and
$\tau_s^{\kappa} \sim \mc{N}(0, \alpha_\eta^{-1})$. 
Now, fix $\hat{\eta}_s = \bs{x}_s^\top \bs{\theta}^\kappa$ and set $\epsilon_{ts} = y_{ts} - \mu_{ts}$.  We then have
$$
\log pr(y_{ts}|\tau_s^{\kappa}, \cdot) = \hat{\eta}_s + \tau^{\kappa}_s - \frac{\xi_s + 1}{\xi_s} \log(h)-h^{-\xi_s^{-1}},
$$
where
$$
h = 1 + \xi_s \epsilon_{ts}\exp(\hat{\eta_s} + \tau_s^{\kappa}).
$$
It follows that
$$
\frac{\partial h}{\partial \tau_s^{\kappa}} = h - 1,
$$
\[
\frac{\partial}{\partial \tau_s^{\kappa}}\log pr(y_{ts}| \tau_s^{\kappa}, \cdot) 
= 1 - \frac{\xi_s + 1}{\xi_s} \frac{h - 1}{h} + \xi_s^{-1}h^{-\xi_s^{-1}} - \xi_s^{-1}h^{-\xi_s^{-1} - 1}
\]
and
\[
\frac{\partial^2}{(\partial \tau_s^{\kappa})^2}\log pr(y_{ts}|\tau_s^{\kappa},\cdot)
= -\frac{\xi_s + 1}{\xi_s} \frac{1}{h^2} - \frac{h^{-\xi^{-1}_s}}{\xi_s^2} + \frac{\xi_s + 2}{\xi_s^2} h^{-\xi_s^{-1} - 1} - \frac{\xi_s + 1}{\xi_s^2} h^{-\xi_s^{-1} - 2}.
\]

\subsection{The case $\xi = 0$}
Using the same notation as above, we have 
\[
\log pr(y_{ts}|\tau_s^{\kappa}, \cdot) = \hat{\eta}_s + \tau_s^{\kappa} + \log(h) - h,
\]
where
\[
h = \exp\left\{-\exp(\hat{\eta}_s + \tau_s^{\kappa}) \epsilon_{ts}\right\}.
\]
Note that 
\[
\frac{\partial h}{\partial \tau_s^{\kappa}} = h \log (h).
\]
Thus
\begin{align*}
\frac{\partial}{\partial \tau_s^{\kappa}} \log pr(y_{ts}|\tau_s^{\kappa},\cdot) &= 1 + \log(h) - h\log(h),\\
\frac{\partial^2}{(\partial \tau_s^{\kappa})^2} \log pr(y_{ts}|\tau_s^{\kappa},\cdot) &= \log(h) - h \log(h)^2 - h\log(h).
\end{align*}

\end{document}